\documentclass[12pt]{article}
\usepackage[top=2.5cm, bottom=3cm, left=2cm, right=2cm]{geometry}
\usepackage{amsfonts}
\usepackage{graphicx}
\usepackage{latexsym,amsmath, color}

\usepackage{natbib}
\usepackage{float}
\begin{document}

\title{Inference with Hamiltonian Sequential Monte Carlo Simulators}
\author{Remi Daviet\thanks{Daviet: Wharton Marketing Department, University of Pennsylvania}}

\date{}
\maketitle

\begin{abstract}
	The paper proposes a new Monte-Carlo simulator combining the advantages of Sequential Monte Carlo simulators and Hamiltonian Monte Carlo simulators. The result is a method that is robust to multimodality and complex shapes to use for inference in presence of difficult likelihoods or target functions. Several examples are provided.
\end{abstract}


\section{Introduction}

Econometrics has traditionally relied on methods based on the optimization of a target function to perform inference. Such methods include the widespread Maximum Likelihood Estimators (MLE), Generalized Method of Moments (GMM) or the Least Squares Estimators (LSE). These methods are especially effective when the target function is unimodal and is shown to converge rapidly to a point. 

Recent developments in economic modeling have lead to the emergence of more complicated target functions where multimodality, complex shapes and slow convergence make traditional inference approaches ineffective. Numerical maximization methods such as Nelder-Mead, Newton-Raphson or Expectation Maximization do not guarantee the convergence toward a global maximum \citep{gourieroux1995statistics}. In addition, these methods provide a point-estimate and do not give information about the shape of the target function.

These more complicated target functions are found in almost every subfield of economics: Multimodal likelihoods for instance can be the result of using DSGE models in Macroeconomics \citep{herbst2014sequential}, GARCH models in Finance \citep{doornik2000multimodality}, BLP models in industrial organization \citep{bajari2003discussion}, Spatial linear models in Urban Economics \citep{mardia1989multimodality}. In fact, with many non-linear models, likelihood functions are non-smooth and multimodal \citep{koop1999bayes}. A similar observation can be made for models with structural breaks and outliers \cite{koop2000nonlinearity}. Various attempts have been made to mitigate the problem, generally relying on Bayesian or Quasi-Bayesian Methods using Markov-Chain Monte Carlo (MCMC) or Sequential Monte Carlo (SMC) simulations. 

MCMC simulation has the advantage of being able to recover more information about a target function than optimization algorithms \citep{chernozhukov2003mcmc}. The only restriction is that the target function has to be positive. In the case of a non-positive target function, the exponential of the function can be taken. This exponentiated function can be treated as a quasi-likelihood. The draws from the simulation provide a Monte Carlo approximation of the distribution of interest, facilitating inference. Application of this method to economics can be found in several papers across all subfields. For instance, Herbst has proposed several Monte Carlo methods to solve the estimation problems with Macroeconomic DSGE models \citep{herbst2014sequential}. We find similar approaches in industrial organization for demand estimation models \citep{jiang2009bayesian}, or in time series analysis \citep{burda2015constrained}. 

The key problem with traditional MCMC simulators using Metropolis-Hastings or Gibbs Sampling methods is that they do not behave necessarily well under multimodality, concentrated mass or complex shapes. The Markov Chain used to simulate the distribution can get trapped in one of the modes if it is not close enough to the other modes and there is no probability mass between them. When facing complex concentrated shapes, these simulators are not performing well exploring the space of interest due to a high level of rejections when trying to move away from the current point in the chain. Some specific simulators have shown interesting properties concerning these problems. The families of Population Monte Carlo (PMC) and more generally SMC simulators solve the problem of multimodality by not having a unique Markov Chain but rather a large set of particles exploring the parameter space \citep{Cappe04populationmonte,delmoral2006sequential,durham2013adaptive}. The sequential approach of the SMC also partly solves the problem of concentrated mass by allowing the target function to progressively converge toward its final form \citep{chopin2002sequential}. The MCMC simulators using Hamiltonian dynamics, sometimes refereed to as Hamiltonian Monte Carlo (HMC) simulators, have shown to be very effective in exploring the parameter space when the target function has a complex elongated shape or isolated concentrated mass. 

Our contribution is to provide a method that is robust to most types of multimodality and complex shapes by combining the advantages of the SMC and the HMC. Moreover, we implement a kernel based resampling method to improve robustness and efficiency. While most econometrics papers describing MCMC or SMC methods are put in a Bayesian framework, we keep the description in a general framework. Using our method, practitioners should be able to perform inference without having to worry about the shape or complexity of their target function.

The advantages of recovering the full shape of the target function are multiple. Multimodality can easily be identified, concentration can be measured and counter-factual checking or prediction can be done by integration over the parameter space. Moreover, the Monte Carlo approach allows for easy variable transformation without the need of derivation of a complicated Jacobian. 

Our method can also be used for maximization of a complicated function $f(x)$ by simulated annealing: the target quasi-likelihood is of the form $\exp(f(x))^\gamma$, where $\gamma$ is chosen by the practitioner. The quasi-likelihood will become concentrated on the maximizers of $f(x)$ as $\gamma \to \infty$ \citep{hwang1980laplace}. The use of SMC methods for optimization via simulated annealing has already been detailed and proven effective \citep{zhou2013sequential}.

We will first present a brief review of the various uses of statistical simulation. We will then describe three common simulations methods (Metropolis-Hastings, HMC, and SMC). We will compare their performance using simple examples. We will finally introduce our HSMC method, its properties and present several simple examples as applications.

\section{Statistical simulation}

Statistical simulation generates samples $\{\theta_n\}_{n=1}^N$ following a distribution with density $f(\theta_n)$ to get an approximation of various quantities depending on that distribution.
These quantities are usually moments, modes or quantiles.

To compute population moments, we can use the sample moments which converge under a Law of Large Number to the desired quantity under weak regularity assumptions. For some continuous function $h(\theta_n)$ and a Euclidean vector $\theta_n$, we have:
$$ \frac{1}{N}\sum_n h(\theta_n) \overset{p}{\to} \int h(\theta) f(\theta) d\theta$$

There are several ways to test for the existence of several modes, count them and find their location. A review of standard kernel based methods can be found in \cite{vieu1996note}. Most popular methods based on critical kernel bandwidth are presented in \cite{silverman1981using} and \cite{minnotte1997nonparametric}. Another popular approach called excess mass approach can be found in several other papers \citep{muller1991excess,butucea2007functional}. The goal of this paper is not to treat mode estimation methods in details and we will leave further readings to the interested reader.

Transformations of the target function can be computed directly by evaluating the transformations of the resulting draws, without the need for analytical derivation or approximation of the Jacobian term which is in many cases very complicated.

\section{Common simulation methods}
As it is often impossible to draw directly from a distribution, several methods have been developed to draw a sample that approximately follows the target distribution. The researcher only needs to be able to evaluate a kernel of the target distribution density function.  

The Metropolis-Hastings (MH) algorithm and the HMC are both Markov Chain Monte Carlo (MCMC) methods. Consider a set of particles $\{\theta_t\}$ whose distribution seeks to approximate an underlying target function of interest $f(x)$. The generic principle of an MH or HMC algorithm is as follows. A particle $\theta_t$ randomly moves to a new position $\theta'_t$ in the space $\Theta$. After each move, with some probability the new position is accepted and $\theta_{t+1}=\theta'_t$. If the move is rejected, $\theta_{t+1}=\theta_t$. The passage from $\theta_t$ to $\theta_{t+1}$ is called an MCMC step. After performing a series of initial steps called burn-in period the Markov Chain should reach its equilibrium distribution which corresponds to the target distribution. We can then use the history of the positions of $\theta_t$ as a sample approximating the target distribution. The difference between MH and HMC is in the way the random move is done, and the formulas to compute acceptance probabilities at the end of the move. 

With SMC methods, $N$ particles  $\{\theta_{nt}\}_{n=1}^N$ are moved simultaneously using MCMC steps and resampled at each iteration. These methods have the advantage of being able to recover the shape of multimodal distributions with separate areas of concentrated mass. These distributions usually cannot be recovered properly using standard MCMC methods as the Markov Chains get trapped under one of the modes without being able to move to the other mode if no probability mass connects them.

\subsection{Metropolis-Hastings}\label{MH}
The Metropolis-Hastings algorithm \citep{metropolis1953equation,hastings1970monte} is the most common accept-reject MCMC method to approximate a target density proportional to $f(\theta)$. Starting from an arbitrary point $\theta_0$, for each iteration a new particle is drawn from a distribution with density $q(\theta'|\theta_t)$. For instance, this distribution could be a $\mathcal{N}(\theta_t,\Sigma)$. Then, we set $\theta_{t+1}=\theta'$ with probability 
$\min(1, \frac{f(\theta')\cdot q(\theta_t|\theta')}{f(\theta_t)\cdot q(\theta'|\theta_t)} )$, and $\theta_{t+1}=\theta_t$ otherwise.

After an initial burn-in period of $b$, the values $\{\theta_t\}_{t=b}^T$ should be approximately distributed folowing the normalized target density.

\subsection{Hamiltonian Monte Carlo}\label{HMC}
As with MH, the purpose of the HMC method is to formulate a Markov chain for which, under certain conditions, a multiple of $f(\theta_t)$ is the density of the stationary distribution. It relies on Hamiltonian dynamics to move a particle $\theta_t$ to a new point $\theta'$ in the $\Theta$ space. This new point, called proposal will then be accepted or rejected as the new value for $\theta_{t+1}$ following a method similar to Metropolis-Hastings rejection step. The movement in the $\Theta$ space can be constrained and we then refer to the method as a Constrained Hamiltonian Monte Carlo (CHMC).

To describe the HMC step, we need to define the function $U(\theta_t)$ with gradient denoted $\nabla U(\theta_t)$:
$$ U(\theta_t) = - \log(f(\theta_t)) $$

We also need to define an auxiliary vector $p$ of dimension $\dim(\theta_t)$ that will represent the momentum of the particle when moving following Hamiltonian dynamics. Intuitively, the particle will move on a surface where the potential energy depending on the altitude is represented by $U_t(\theta_n)$. When the particle goes up a slope, it will slow down or even turn back. The continuous movement is approximated by a series of $L$ steps of size $\epsilon$. At the end of the last step, the momentum is reversed to make the proposal symmetric. If we start at the final position with the final reversed momentum, we will find the particle going back to the original position after $L$ steps. This ensures reversibility and facilitate the computation of the acceptance probability.

The HMC step proceeds as follow:

\begin{enumerate}
	\item define the starting position of the proposal $\theta' = \hat{\theta}_t$
	\item draw an initial momentum vector $p$ from a multivariate Gaussian distribution: $p \sim \mathcal{N}(0,M)$
	\item update the momentum vector by half a step taking the gradient into account: $p'= p - \frac{\epsilon}{2} \cdot \nabla U(\theta')$
	\item Repeat for $l = 1,\ldots,L$
	\begin{enumerate}
		\item \label{posStep} update the position by a full step: $\theta' = \theta' + \epsilon \cdot p' $
		\item update the momentum by a full step, except at the end of the trajectory: if $(l\neq L)$, then  $p' = p' - \epsilon \cdot \nabla U(\theta')$
	\end{enumerate}
	\item update the momentum vector by half a step: $p'= p' - \frac{\epsilon}{2} \cdot \nabla U(\theta')$
	\item negate the momentum vector: $p' = - p'$
	\item compute the acceptance probability:\\ $a = \min\left(1, \exp\left[U(\theta_t)-U(\theta')+\frac{\sum p^2}{2} -\frac{\sum {p'}^2}{2}\right] \right)$
	\item set $\theta_{t+1}=\theta'$ with probability $a$, and $\theta_{t+1}=\theta_t$ otherwise
\end{enumerate}

The HMC mutation step needs to be tuned by choosing appropriately the quantities $(M,L,\epsilon)$. To learn more about how to choose these quantities, see the review paper by \cite{neal2011mcmc}.

When we want to put constraints on some of the dimensions of $\theta$, we can modify the HSMC method to have the particles to bounce off the constraints as if they were walls. Most of the time, the constraints are of the form $\theta_{dt} \leq u_d$ or $\theta_{dt} \geq l_d$ for some dimensions $d$ of $\theta_t$. The position updating step \ref{posStep} is then replaced by:
\begin{enumerate}
	\item for each dimension $d$ of $\theta'$:
	\begin{enumerate}
		\item update position in dimension $d$: $\theta'_d = \theta'_d + \epsilon \cdot p'_d $
		\item if $\theta'_d$ is constrained, repeat the following until $\theta'_d$ satisfies all constraints:
		\begin{enumerate}
			\item if $(\theta'_d>u_d)$, then $\theta'_d=u_d - (\theta'_d-u_d)$ and $p'_d=-p'_d$
			\item if $(\theta'_d < l_d)$, then $\theta'_d=l_d + (l_d - \theta'_d)$ and $p'_d=-p'_d$
		\end{enumerate}
	\end{enumerate}
\end{enumerate}

With this modified approach, if the particle passes a constraint "wall" during a position update, the symmetric of the particle $\theta'$ relative to the wall in the $\Theta$ space is taken, and the momentum in the constrained dimension is reversed. This approach simulates the particle bouncing off the wall and preserves reversibility.
A similar approach can be used for more complex constraints of the form $G(\theta)\geq 0$.

\subsection{Sequential Monte Carlo}
As mentioned before, SMC methods use multiple particles moving in parallel \citep{delmoral2006sequential}. While there are many different types of SMC, we are going to describe one of the most popular approaches alternating Importance Resampling and MH steps. The goal is to obtain a sample $\{\theta_n\}_{n=1}^N$ from a sequence of distributions  with densities $f_1(\theta_n),\ldots,f_T(\theta_n)$. The simulator we propose works best when target densities in the sequence are smoother at the beginning and progressively converging toward the final sharper target density. We also require an initial distribution with density $f_0(\theta_n)$ that is easy to sample from and covers well the mass of the first distribution in the sequence. An SMC simluator provides us for each step $t=1,\cdots,T$  with a set of values $\{\theta_n\}_{n=1}^N$ that approximately follow the distribution $f_t(\theta_n)$.

A sequence of distributions can be found in many applications relevant to economics. In frequentist econometrics, the sequence can be the likelihood or quasi-likelihood for data collected until time $t$, e.g. $f_t(\theta_n) \propto L(\theta_n;y_1,\ldots,y_t)$. The quasi-likelihood can be built from any estimator maximizing or minimizing a target function such as the Generalized Method of Moments or a Least Squares Estimator \citep{chernozhukov2003mcmc}. The counterpart in the Bayesian framework would be the posterior distribution of the parameter $\theta$ given the data until time $T$. Compared to a standard MCMC approach which requires an evaluation of the target function with every observation at each step, the SMC approach is less computationally intensive. Moreover, adding the observations one or a few at a time creates a desirable tempering effect \citep{chopin2002sequential}. This approach is particularly efficient in large datasets where new observations come regularly as updating the estimator can be done in one step. 

A second application is kernel density estimation when observations are added a few at a time and bandwidth is progressively shrunk. An application of this approach can be found later in this paper. 

Another application proposed by \cite{neal2001annealed} shows the benefits of moving progressively from a tractable distribution $f_1(\theta_n)$ to a target distribution $f(\theta_n)$ by geometrically reweighting them : $ f_t(\theta_n) \propto f(\theta_n)^{\phi_t} f_1(\theta_n)^{1-\phi_t}$ with $0 \leq \phi_1 < \ldots < \phi_T = 1$.

Finally, SMC can be used for maximization using a simulated annealing approach. Our sequence of distributions will then become $ f_t(\theta_n) \propto f(\theta_n)^{\gamma_t}$ with $\gamma_t$ increasing to high values as $t$ increases.

The algorithm is as follow:

\begin{enumerate}
	\item Initialization: Draw $N$ particles $\{\theta_n^{(0)}\}_{n=1}^N$ from $f_0(\theta_n)$
	\item Repeat for $t =1,\ldots,T$
	\begin{enumerate}
		\item \label{algo:corr} Correction: assign weight $w_n^{(t)} = f_{t}(\theta_n) / f_{t-1}(\theta_n)$ to each of the particles $\{\theta_n^{(t-1)}\}_{n=1}^N$
		\item \label{algo:sel} Selection: draw $N$ new particles $\{\hat{\theta}_n^{(t)}\}_{n=1}^N$ with replacement from the current sample of particles using weights $w_n^{(t)} $. Give the new particles a weight of $1$.
		\item Mutation: For each particle, perform a MH step as described in section \ref{MH} to obtain a new sample of particles $\{\theta_n^{(t)}\}_{n=1}^N$.
	\end{enumerate}
\end{enumerate}

\subsection{Examples}

We are now going to illustrate the advantages of various algorithms with simple examples. The first example is a Rosenbrock's banana function and illustrates the problems classical MH face when the function has an elongated shape. The second example is a 6-dimensional normal distribution. This example is designed to show the difference in convergence speeds when dimensionality increases.

\subsubsection{Banana function}

\begin{figure}[h]
	\caption{Contour plot of the Rosenbrock function}
	\label{fig:rosenbrock}
	\begin{center}
		\includegraphics[scale=0.4]{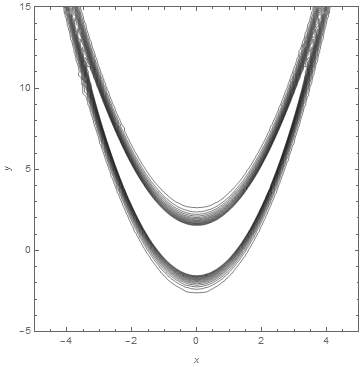}
	\end{center}
\end{figure}

The Rosenbrock's banana function (Figure \ref{fig:rosenbrock}) is defined as follow: $$f(\theta) \propto e^{\frac{1}{8} \left(-5 \left(y-x^2\right)^2-x^2\right)}$$ 

\begin{figure}[H]
	\caption{ Metropolis-Hastings paths for a Rosenbrock function}
	\label{fig:MH-Rosen}
	\begin{center}	
		\includegraphics[scale=0.3]{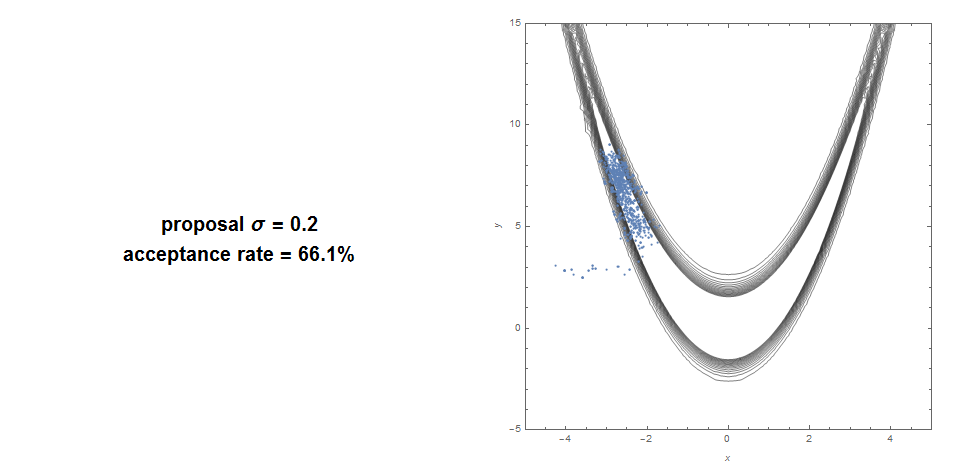}\\
		\includegraphics[scale=0.3]{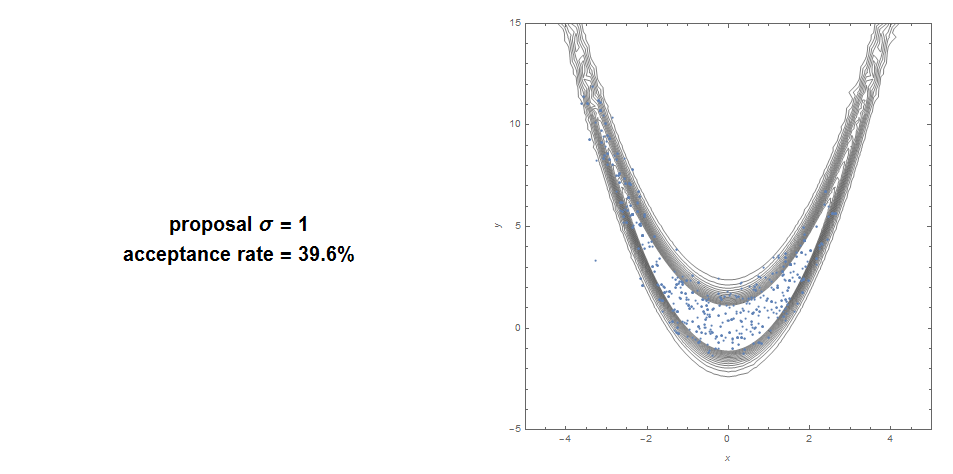}
	\end{center}
\end{figure}

Using the MH algorithm, we run two trials with different $\sigma$ for the $\mathcal{N}(\theta,\sigma I)$ proposal. When $\sigma=0.2$, the acceptance rate is $66.1\%$ but the chain fails to cover the distribution after 1000 iterations. When $\sigma=1$, the coverage improves but the acceptance rate decreases to $39.6\%$. A visual representation of the paths is provided in Figure \ref{fig:MH-Rosen}. 

\begin{figure}[H]
\caption{ Hamiltonian Monte-Carlo path for a Rosenbrock function}
\label{fig:HMC-Rosen}
\begin{center}	
	\includegraphics[scale=0.3]{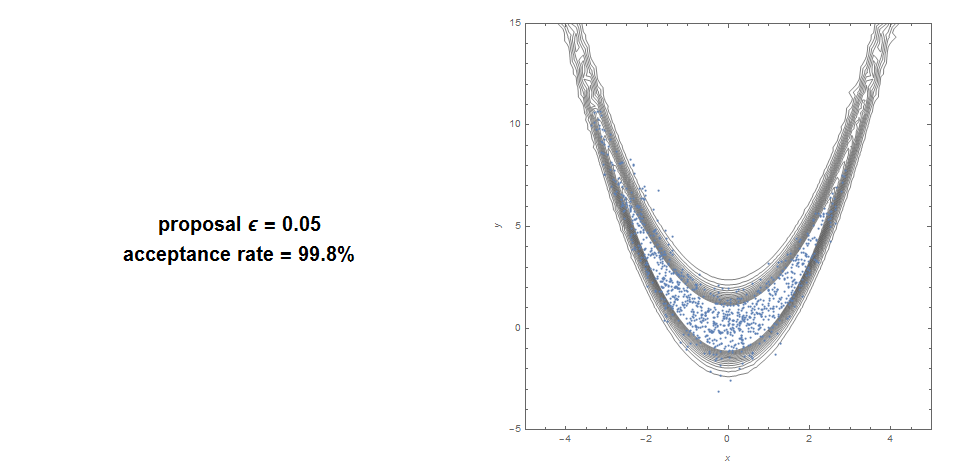}
\end{center}
\end{figure}

Using the HMC approach, with $\epsilon = 0.05$ and $L=20$ steps, the target density is well covered and the acceptance rate is $99.8\%$. A visual representation of the paths is provided in Figure \ref{fig:HMC-Rosen}.

\subsubsection{6-dimensional normal distribution}

Using a $N(\mu,I)$ as the target, where $\mu=(10,10,10,-10,-10,-10)$, we compare the performance of the MH and HMC approach starting the chain at the point $(-15,-15,-15,15,15,15)$. The $\sigma$ for the MH method is chosen to get at least a $60\%$ acceptance rate.

\begin{figure}[H]
	\caption{Comparison of Metropolis-Hastings path (Top) and Hamiltonian Monte-Carlo path (Bottom) for a 6-dimensional normal distribution.}
	\label{fig:gaussian-6-dim}
	\begin{center}
		\includegraphics[scale=0.3]{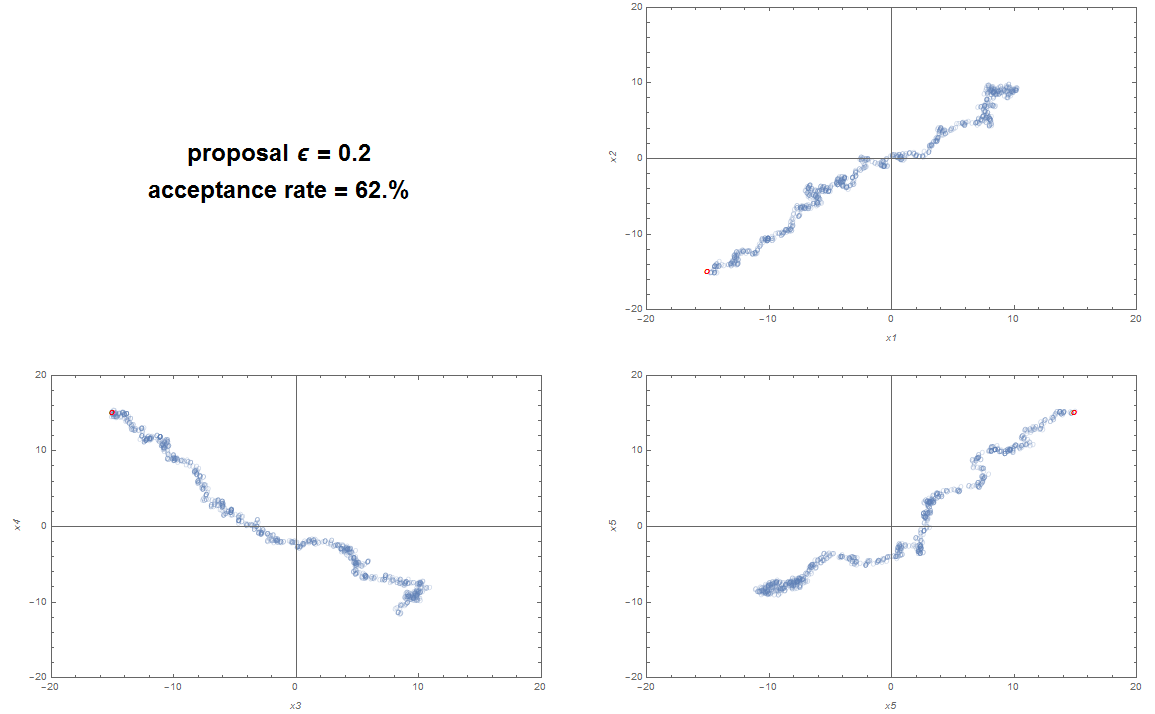}
		\includegraphics[scale=0.3]{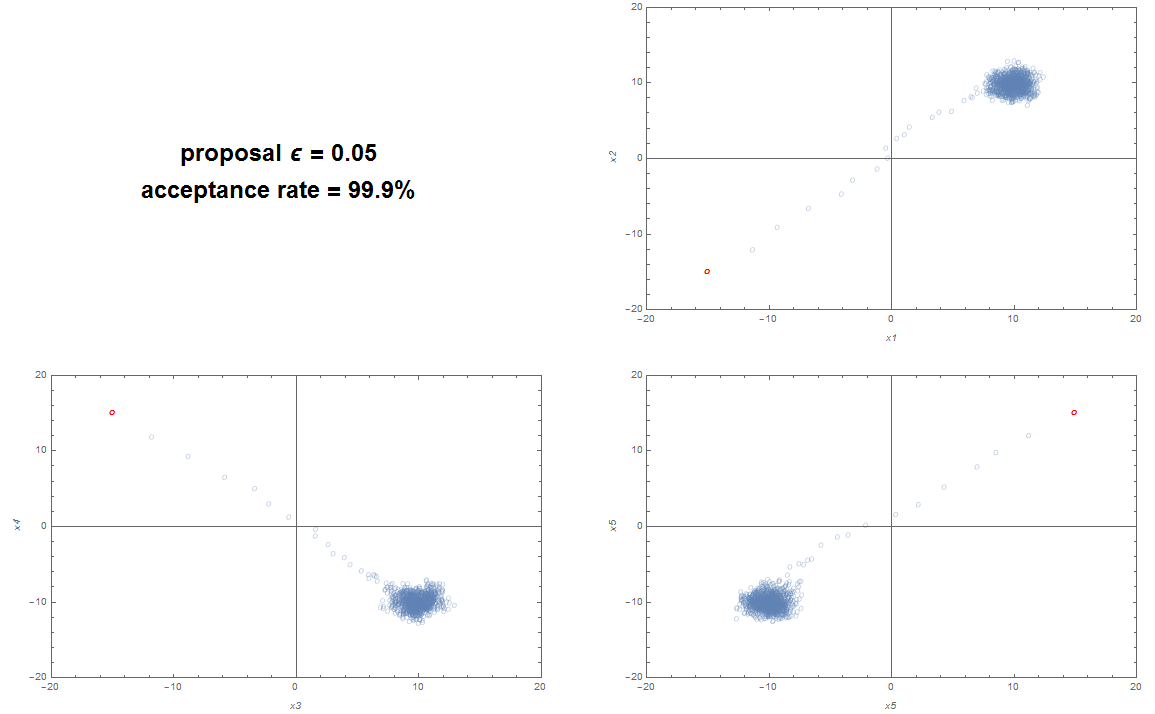}
	\end{center}
\end{figure}

In the case of the MH approach, the chain requires a burn in of about 800 iterations to reach the mass of the target density. With the HMC approach, only 50 iterations are required for the chain to converge. A visual representation of the paths is provided in Figure \ref{fig:gaussian-6-dim}.

\section{Hamiltonian Sequential Monte Carlo}

In this section, we are going to describe our method and its properties. The HSMC method replaces the MH step of standard SMC by a Hamiltonian step. The re-sampling is also done using a leave-one-out approximation $\hat{f}_{t-1}(\theta_n)$ of the observed distribution of the particles instead of the theoretical distribution $f_{t-1}(\theta_n)$ as in specific cases the particles do not have the time to converge to their stationary distribution in one step. This method shows good convergence rates when applied to importance resampling \citep{delyon2016integral}.

\subsection{Algorithm}

\begin{enumerate}
	\item Initialization: Draw $N$ particles $\{\theta_n^{(0)}\}_{n=1}^N$ from $f_0(\theta_n)$
	\item Repeat for $t =1,\ldots,T$
	\begin{enumerate}
		\item \label{algo:corr} Correction: assign weight $w_n^{(t)} = f_{t}(\theta_n) / \hat{f}_{t-1}(\theta_n)$ to each of the particles $\{\theta_n^{(t-1)}\}_{n=1}^N$, where $\hat{f}_{t-1}(\theta_n)$ is a "leave-one-out" kernel density estimate.
		\item \label{algo:sel} Selection: draw $N$ new particles $\{\hat{\theta}_n^{(t)}\}_{n=1}^N$ with replacement from the current sample of particles using weights $w_n^{(t)} $. Give the new particles a weight of $1$.
		\item Mutation: For each particle, perform a Hamiltonian step as described in section \ref{HMC} to obtain a new sample of particles $\{\theta_n^{(t)}\}_{n=1}^N$.
	\end{enumerate}
\end{enumerate}

In the initialization phase we obtain a sample of particles distributed according to $f_0(\theta_n)$ distribution. For the HSMC method to perform well we need a distribution that covers well the whole $\Theta$ space and has mass where the other distributions $f_t(\theta_n)$ in the sequence have mass too.

In the loop, before the correction phase, we have particles all weighted to $1$ that provides a Monte-Carlo simulation of the distribution $f_{t-1}(\theta_n)$. In the correction phase, we reweight them to obtain an approximation of the distribution $f_t(\theta_n)$ by importance sampling.

In the selection phase, we perform sampling importance resampling to obtain an approximation of the distribution $f_t(\theta_n)$ using particles $\{\hat{\theta}_n^{(t)}\}_{n=1}^N$ with equal weights. Note that at the end of the selection phase, we can expect to have several particles sharing the same value.

Finally, in the mutation phase, we explore the $\Theta$ space by moving the particles using a Hamiltonian Monte Carlo (HMC) approach. Since $f_t(\theta_n)$ is the stationary distribution of our HMC, the particles both before and after the HMC step should approximate $f_t(\theta_n)$. However, in the case $\{\hat{\theta}_n^{(t)}\}_{n=1}^N$ do not follow exactly the distribution $f_t(\theta_n)$, performing a HMC step should improve the approximation by the convergence properties of HMC \citep{neal2011mcmc}.

The main contribution of the paper lies in the use of Hamiltonian dynamics for the mutation phase in a SMC method. In the literature, SMC algorithms are generally found to use a standard Metropolis-Hastings step in their mutation phase. Conversely, HMC methods using multiple particles do not have the resampling phases \ref{algo:corr}/\ref{algo:sel}.

\subsection{Algorithm's properties}
Our HSMC method fits into the SMC framework described in \cite{chopin2004central} and his central limit theorem can be applied to our simulator. Provided that our Hamiltonian mutation step preserves the distribution $f_t(\theta_n)$, the following convergences hold almost surely as $N \to \infty$ for any measurable function $\phi$ such that the expectations below exists:
$$ N^{-1} \sum_{n=1}^{N} \phi(\theta_n^{(t)}) \to \mathbb{E}_{f_{t}}[\phi(\theta_n)] $$
$$ \frac{\sum_{n=1}^{N} w_n^{(t)} \phi(\theta_n^{(t-1)})}{\sum_{n=1}^{N}w_n^{(t)}} \to \mathbb{E}_{f_{t}}[\phi(\theta_n)] $$
$$ N^{-1} \sum_{n=1}^{N} \phi(\hat{\theta}_n^{(t)}) \to \mathbb{E}_{f_{t}}[\phi(\theta_n)] $$

The proof that our Hamiltonian transition kernel satisfies the conditions to have $f_t(\theta_n)$ as a stationary distribution can be found in the review paper on Hamiltonian Monte Carlo by \cite{neal2011mcmc}.

\section{Method variations}
While the Hamiltonian step is designed to be applied on continuous distributions, it is easy to extend our method for spaces with discrete dimensions. We can split our space $\Theta$ in two blocks $\{\Theta_c,\Theta_d\}$ where $\Theta_c$ includes the dimensions where $\Theta$ is continuous and $\Theta_d$ includes the dimensions where $\Theta$ is discrete. From this separation into blocks, a standard Metropolis within Gibbs \citep{gilks1995adaptive} step can be used with the Hamiltonian step being used in the continuous block. 

Another variation of the method consists in running the algorithm in parallel for $J$ groups of $N$ particles. At the end, the sample properties of the $J$ groups can be compared. If the properties differ too much from each other, we can suspect a convergence problem. This is the approach taken by \cite{durham2013adaptive} in their adaptive SMC simulator.

Finally, since the Hamiltonian step preserves the distribution of interest, multiple steps can be made at each mutation phase. This solution increases the performance of the method when particles are not exploring the $\Theta$ space fast enough.

\section{Examples}
We used the HSMC method on kernel density estimates of two functions known to challenge classical optimizers and MCMC simulators. The first function is created for this paper and called the smiley function. It is a mixture of 3 Rosenbrock smile functions often found as an example to show the limitations of MCMC algorithms. The second function is a dropwave function. Both functions present multimodality and follow complex shapes. For ease of visualization, we kept the functions two-dimentional.

We chose to simulate Gaussian kernel density estimates as they can mimic the progressive convergence of several target functions when data are added progressively. In our case, data are added by blocks of $100$ points and the bandwidth of the kernel density estimate is $n^{-1/5}$. This rate has been chosen as it is the order of the bandwidth reduction rate when using optimal bandwidth for most distributions \citep{yatchew1998nonparametric}.

We also used the algorithm on a non-linear logit model where the log-likelihood shows both multimodality and complex shapes.

\subsection{Smiley kernel density estimate}

\begin{figure}[H]
	\caption{contour plot and 3D plot of the smiley function}
	\label{fig:smileplots}
	\begin{center}
	\includegraphics[scale=0.5]{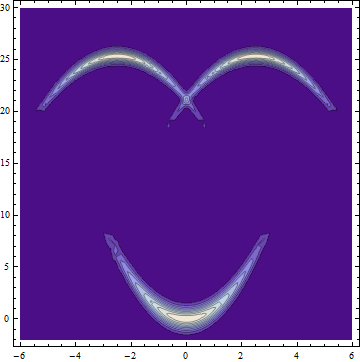}
	\includegraphics[scale=0.55]{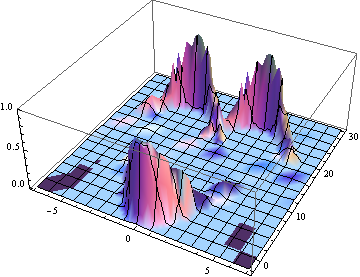}
	\end{center}
\end{figure}

\begin{figure}[!ht]
	\caption{Generated data}
	\label{fig:smiledata}
	\begin{center}
	\includegraphics[scale=0.55]{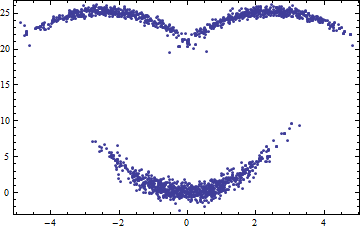}
	\end{center}
\end{figure}

We generated a sample of 2048 data points with coordinates $(x,y)$ using a density proportional to the following function:
\begin{align*}
g(x,y) = & \exp \left(\frac{1}{5} \left(-6 \left(-(2.5\, -x)^2-1.5 y+38\right)^2-(2.5\,
-x)^2\right)\right) \\
 &+\exp \left(\frac{1}{5} \left(-6 \left(-(x+2.5)^2-1.5
y+38\right)^2-(x+2.5)^2\right)\right) \\
&+\exp \left(\frac{1}{5}  \left(-5
\left(y-x^2\right)^2-x^2\right)\right)
\end{align*}
The contour plot and 3D plot of the function used can be found in Figure \ref{fig:smileplots}.
We can easily see the multimodality and the elongated shapes on the function plots. The data generated match the shape of the smiley function and have been represented in Figure \ref{fig:smiledata}.

\begin{figure}[H]
	\caption{HSMC Simulation results}
	\label{fig:smilesim}
	\begin{center}
		\includegraphics[scale=0.5]{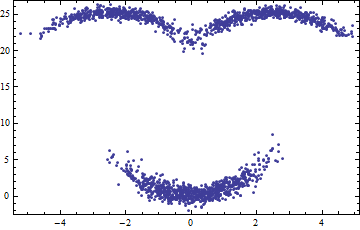}
		\includegraphics[scale=0.5]{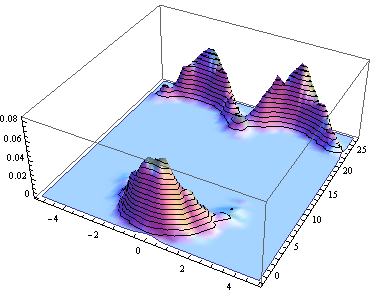}
	\end{center}
\end{figure}

The target function we want to simulate is a kernel density estimate using these data points. We used 4 independent groups of 512 particles to simulate the kernel density for a total of 2048 particles. The data points are partitioned in 20 blocks of 100 points and 1 block of 48 points, for a total of 21 blocks. Consequently, the sequence will include $T=21$ target functions with the addition of the initial density. The initial density $f_0(\theta_n)$ used is a bivariate normal distribution with parameters $\{\mu_1=0,\mu_2=10,\sigma_1=10,\sigma_2=20,\rho=0\}$. The HMC tunning parameters $(M,L,\epsilon)$ used are $(I_2,20,0.05)$, where $I_2$ is the $2\times 2$ identity matrix. The results of the HSMC simulation as well as a smoothed histogram of the simulated points have been represented in Figure \ref{fig:smilesim}. We can see that the HSMC method provided satisfying results and successfully converged with only 21 iterations. The lowest mutations acceptance rate we observed across several runs was $2043/2048$.

\begin{figure}[H]
	\caption{Parallel HMC Simulation results}
	\label{fig:smilespara}
	\begin{center}
		\includegraphics[scale=0.5]{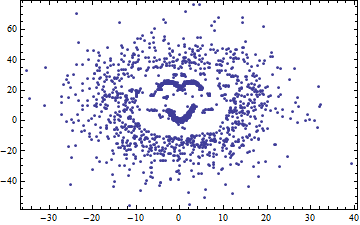}
	\end{center}
\end{figure}

To illustrate the performance of our HSMC method, we also tried to simulate the same sequence of functions using a parallel HMC approach with 21 iterations on the same kernel density. To do so we performed the mutation phase 21 times on $f_{21}(\theta_n)$. The result represented in Figure \ref{fig:smilespara} show that several particles failed to converge. Moreover, there is no guarantee that the particles around each mode are distributed according to the mass around these modes in the target function. As particles tend to converge to the closest mode, it is possible to have a first mode with twice the mass of a second mode but only half of the particles around it.

\subsection{Constrained dropwave kernel density estimate}

\begin{figure}[H]
	\caption{contour plot and 3D plot of the dropwave function}
	\label{fig:dropplots}
	\begin{center}
	\includegraphics[scale=0.5]{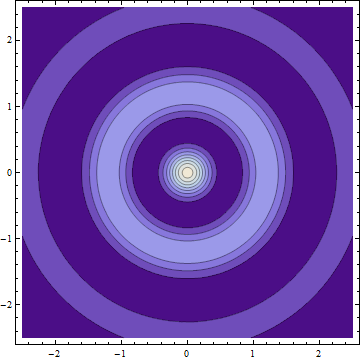}
	\includegraphics[scale=0.55]{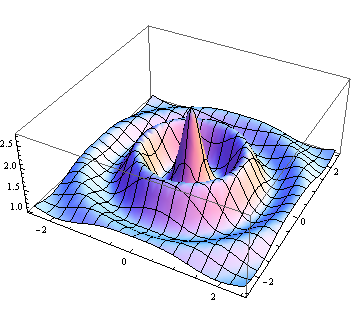}
	\end{center}
\end{figure}

We generated a sample of 4096 data points with coordinates $(x,y)$ using a density proportional to the following function defined on $[-2.5;2.5]\times[-2.5;2.5]$:
\begin{align*}
g(x,y) = & \exp\left(\frac{\cos \left(5 \sqrt{x^2+y^2}\right)+1}{x^2+y^2+2}\right)
\end{align*}
The contour plot and 3D plot of the function used can be found in Figure \ref{fig:dropplots}.

\begin{figure}[H]
	\caption{Generated data}
	\label{fig:dropdata}
	\begin{center}
		\includegraphics[scale=0.55]{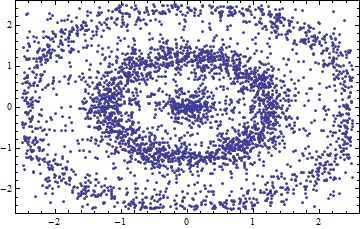}
	\end{center}
\end{figure}

The data generated match the shape of the dropwave function and have been represented in Figure \ref{fig:dropdata}.

\begin{figure}[H]
	\caption{CHSMC Simulation results}
	\label{fig:dropsim}
\begin{center}
		\includegraphics[scale=0.5]{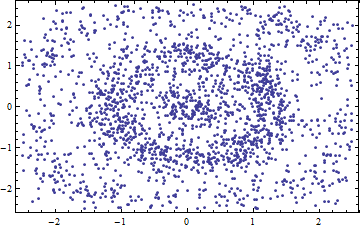}
	\includegraphics[scale=0.5]{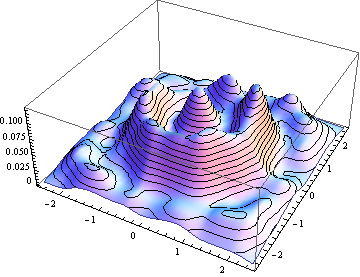}
\end{center}
\end{figure}

The target function we want to simulate is a constrained kernel density estimate using these data points. The Gaussian kernel is defined on $\Bbb{R}^2$ but we want to limit the domain to $[-2.5;2.5]\times[-2.5;2.5]$. To do so we keep the Gaussian kernel as is but use the constrained HMC method during our mutation phase.  We used 4 independent groups of 512 particles to simulate the kernel density for a total of 2048 particles. The data points are partitioned in 40 blocks of 100 points and 1 block of 96 points, for a total of 41 blocks. Consequently, the sequence will include $T=41$ target functions with the addition of the initial density. The initial density $f_0(\theta_n)$ used is a bivariate normal with parameters $\{\mu_1=0,\mu_2=0,\sigma_1=10,\sigma_2=10,\rho=0\}$. The HMC tunning parameters $(M,L,\epsilon)$ used are $(I_2,20,0.05)$, where $I_2$ is the $2\times 2$ identity matrix. The results of the HSMC simulation as well as a smoothed histogram of the simulated points are presented in Figure \ref{fig:dropsim}. We can see that the constrained HSMC method also provided satisfying results and successfully converged with 41 iterations. The lowest mutations acceptance rate we observed across several runs was $2023/2048$.

\subsection{Non-linear logit}

\begin{figure}
	\caption{$V(x)$, $x \in [-2;8]$}
	\label{fig:nonlinutil}
	\begin{center}
		\includegraphics[scale=0.5]{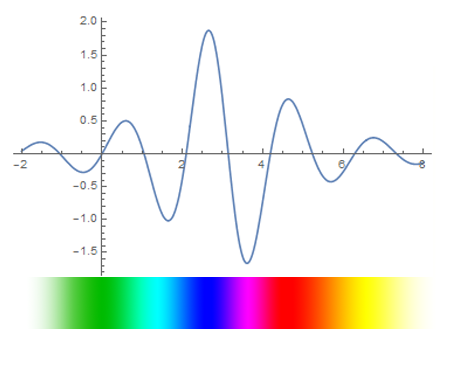}
	\end{center}
\end{figure}

We use a Logit discrete choice model \citep{train2009discrete} where the deterministic part of the utility function is non-linear. In this hypothetical experiment, an individual who has been given a black t-shirt is presented with another t-shirt of a different color $x_t$. The individual can exchange it against his own t-shirt $x_0$ or keep his current t-shirt.
The individual chooses $x_t$ or $x_0$ in order to maximize utility represented by:
$$ U(x_t) = V(x_t) + e_{1t} $$
$$ U(x_0) = e_{0t}$$
where $e_{1t}$ and $e_{0t}$ are the standard extreme value errors. The deterministic part of the utility function is represented by the fuction over the color palette $x \in [-2;8]$:
$$ V(x) = 2 \frac{\sin(\beta_2 \cdot x)}{1+0.5 \cdot (\beta_1 - x)^2} $$
This function is represented on Figure \ref{fig:nonlinutil} for the values $beta_1=3$ and $beta_2=3$.

The choice probabilities for $x_t$ and $x_0$ can be shown to equal:
$$ \frac{e^{V(x_t)}}{1+e^{V(x_t)}} \qquad , \qquad 1- \frac{e^{V(x_t)}}{1+e^{V(x_t)}} $$
We can note that as $\beta_1 \to \infty$, the choice probabilities converge to $0.5$.
We simulated $400$ experiments by drawing randomly an $x_t$ uniformly on $[-2;8]$.
\begin{figure}[H]
	\caption{Theoretical Likelihood}\label{fig:nonlinlik}
	\begin{center}
		\includegraphics[scale=0.5]{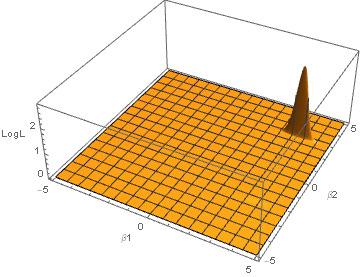}
	\end{center}
\end{figure}
\begin{figure}[H]
	\caption{Theoretical Loglikelihood}\label{fig:nonlinloglik}
	\begin{center}
			\includegraphics[scale=0.5]{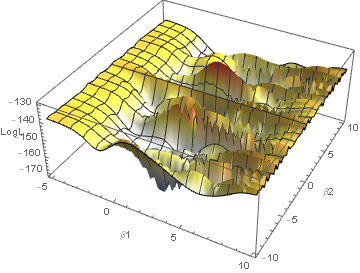}
			\includegraphics[scale=0.4]{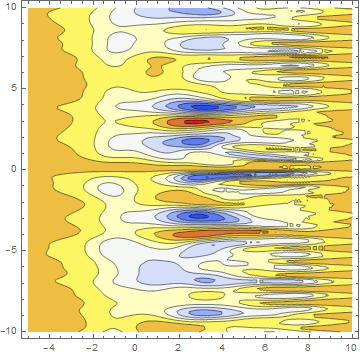}
	\end{center}
\end{figure}

A first look at the shape of the likelihood function represented in Figure \ref{fig:nonlinlik} seems to indicate that the likelihood behaves nicely with a global maximum around the true value. However, examining the log-likelihood on Figure \ref{fig:nonlinloglik} reveals the multimodality of the function.

The MH and HMC methods both fail ton converge after $5000$ iterrations and do not get close to the global-maximum on numerous trials.

For the HMC method, the sequence of functions is created by adding the observations progressively in the likelihood function, adding $50$ observations for each iteration. We compare the result to a standard SMC algorithm.

\begin{figure}[H]
	\centering
	\begin{minipage}{0.5\textwidth}
		\centering
		\includegraphics[scale=0.4]{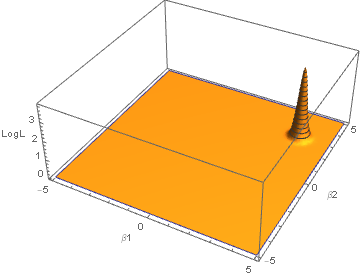}
		\caption{Smoothed HSMC simulation} \label{fig:nonlin-HSMC}
	\end{minipage}%
	\begin{minipage}{0.5\textwidth}
		\centering
		\includegraphics[scale=0.4]{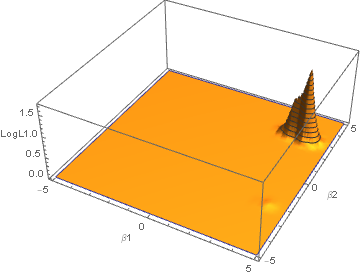}
		\caption{Smoothed SMC simulation} \label{fig:nonlin-SMC}
	\end{minipage}
\end{figure}

We can see that the HSMC method (Figure \ref{fig:nonlin-HSMC}) could converge in only 4 iterations, whereas the SMC method (Figure \ref{fig:nonlin-SMC}) couldn't converge completely. First, in the SMC case, $4$ iterations might constitute a too short sequence to obtain convergence. Adding more intermediate functions in the sequence might improve convergence. Second, we notice that some of the particles in the SMC case were trapped in a mode around the point $(4,-2)$ and these particles are unlikely to be moved to the main area of mass concentration with more iterations. With the HSMC algorithm, if some particles are trapped in a different mode and misrepresent the mass in the underlying function, they get relocated during the resampling phase. This is the consequence of using  $f_{t}(\theta_n) / \hat{f}_{t-1}(\theta_n)$ as a weight instead of the theoretical $f_{t}(\theta_n) / f_{t-1}(\theta_n)$ used in standard SMC.

\section{Future work and conclusion}

We have shown that the HSMC algorithm is able to approximate by simulation many functions with irregularities.
The encouraging performance of the algorithm finds direct potential applications in empirical work where standard MCMC methods have shown limitations. One of this potential application is the use of optimal instruments for non-linear BLP models in industrial organization \citep{reynaert2014improving}.

\bibliography{HSMC}

\end{document}